\documentclass[doublecol]{epl2}
\input epsf.sty
\usepackage{amsmath}
\usepackage{graphicx}
\usepackage{dcolumn}
\usepackage{bm}

\title{Formation of nonequilibrium modulated phases under local energy input}
\shorttitle{Formation of nonequilibrium modulated phases under local energy input}
\author{Linjun Li \and Michel Pleimling}
\shortauthor{Linjun Li and Michel Pleimling}
\institute{
\inst{1} Department of Physics, Virginia Tech, Blacksburg, VA 24061-0435 USA}

\abstract{
We study numerically an inhomogeneous Ising lattice gas with short-range interactions where different sectors are in contact with thermal baths at different
temperatures. Inside the different sectors particles jump to empty sites following the familiar Kawasaki dynamics. In addition,
particles can freely hop from one sector to the other. This crossing between the sectors breaks detailed balance and
yields a local energy influx that 
drives the system to a nonequilibrium steady state. 
When the low-temperature sector is cooled below the equilibrium critical temperature, a complicated nonequilibrium phase diagram
emerges, dominated by unusual modulated nonequilibrium stationary states. These steady states result from the interplay
of phase separation and convection.}

\begin{document}
\maketitle

Systems far from equilibrium display a large variety of novel and unexpected features that often defy our common sense.
Enhancing our understanding of generic properties of systems far from equilibrium therefore remains one of the main challenges
faced by contemporary physics. Recent years witnessed some major progress in the theoretical studies of various typical nonequilibrium
situations, encompassing steady-state properties of paradigmatic transport models \cite{Cho11} and driven diffusive systems \cite{Sch95},
aging phenomena during relaxation processes \cite{Hen10} as well as fluctuation relations and theorems both for steady-state systems
and for systems driven out of a steady state (see, e.g., \cite{Eva93,Gal95,Leb99,Jar97,Cro99,Har07}).
Still, a common theoretical framework for nonequilibrium systems remains elusive.

The role of surfaces and interfaces during nonequilibrium processes remains poorly understood as the
overwhelming majority of studies focuses on bulk systems (see 
\cite{Jan88, Har88, Fro01, Ple04a, Ple04b, Ple05, Bau07, Kad08,Huc09, Igl11, Agl07, Agl11} for some
examples where the effect of surfaces was discussed).
However, as long-range correlations are usually present
far from equilibrium, and this even in systems with only short-range interactions, the influence of surfaces and
interfaces is expected to be non negligible in many instances, thereby changing the physical properties even far away from the
interface.

In this Letter we discuss the intriguing phase diagram that emerges in an interacting many-body
system with conserved dynamics when energy is
pumped into the system locally at an interface. This nonequilibrium phase diagram is found to be dominated by modulated phases 
with a modulation vector parallel to the interface.

\begin{figure} [h]
\includegraphics[width=0.90\columnwidth]{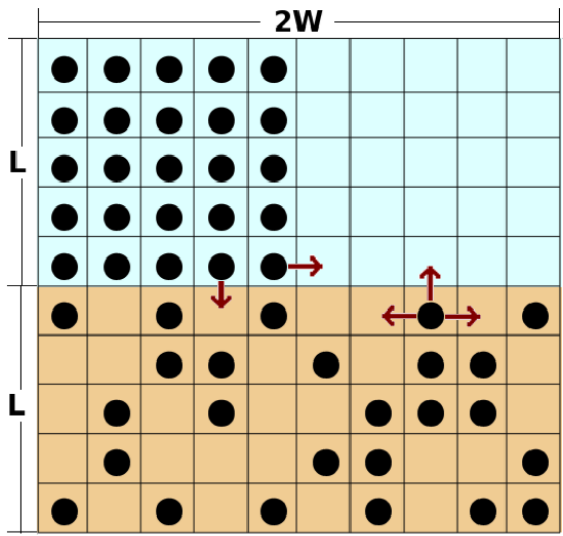}
\caption{\label{fig1} (Color online)
Schematic sketch of our system. The lower (upper) sector is in contact with a hot (cold) thermal bath.
Disorder prevails in the hot sector, whereas phase separation is possible 
in the cold sector when cooled below the critical temperature of the two-dimensional Ising system.
The arrows indicate possible moves of some particles. Note that the particles can freely jump
from one sector to the other, thereby breaking detailed balance.}
\end{figure}

The model we consider in the following is the standard ferromagnetic two-dimensional Ising model with the Hamiltonian
      \begin{equation}
      \label{eq:Ham2dIsing}
      H=-J \sum_{x,y}{(S_{x,y}S_{x+1,y}+S_{x,y}S_{x,y+1})}
      \end{equation}
where the spin $S_{x,y} = \pm 1$ characterizes the state of the site $(x,y)$, whereas $J > 0$ is the coupling constant
between nearest neighbor spins. 
We supplement this model with conserved dynamics that only allows spin exchanges.
Alternatively, we can cast our model in the language of the Ising lattice gas where the spin value $+1$ corresponds to
a particle, whereas the value $-1$ indicates an empty site (hole). 
The total magnetization $M$ of our system (or, equivalently, the particle density) is therefore kept fixed.
In this communication we restrict ourselves to the case $M = 0$.
This system is brought into contact with two different
thermal baths, see Fig. \ref{fig1}: whereas the lower sector is immersed into a heat bath that is at some high temperature $T'$
(the data presented below have been obtained for $T' = \infty$, for the sake of simplicity), the upper sector is in
contact with a much colder heat bath, with temperature $T_0$. Using periodic boundary conditions in both directions,
our systems consist of $2 W \times 2 L$ sites, with two interfaces separating the hot and cold sectors. 
Inside the two sectors we use Kawasaki dynamics, so that
a particle can jump to an empty neighboring site (spins of different signs on neighboring sites can be exchanged) with the Metropolis rates at
the given temperature. In addition, particles freely jump from one
sector to the other. 
This breaks detailed balance at the interface separating the sectors and results
in an energy flow from the hot sector to the cold one. As a result the system is driven to
a nonequilibrium stationary state \cite{Ple10}.

We carefully checked that the results of our Monte Carlo simulations reported in the following were indeed obtained in the steady state of our
systems. Starting from different initial conditions, the system settles in the same stationary state after a long transient.
For the half width $W = 60$ discussed in detail below we typically discard the first 7 million time steps (one time step corresponds
to $2W \times 2L$ proposed updates) before computing time averages. Longer relaxation times are needed for larger system sizes.
Our data acquisition runs were started from an initial state with half of the particles randomly distributed
in the lower sector (corresponding to an equilibrium state at infinite temperature), whereas in the upper sector we considered
a fully phase separated configuration (corresponding to the $T=0$ equilibrium state), see Fig. \ref{fig1}.

\begin{figure} [h]
\includegraphics[width=0.95\columnwidth]{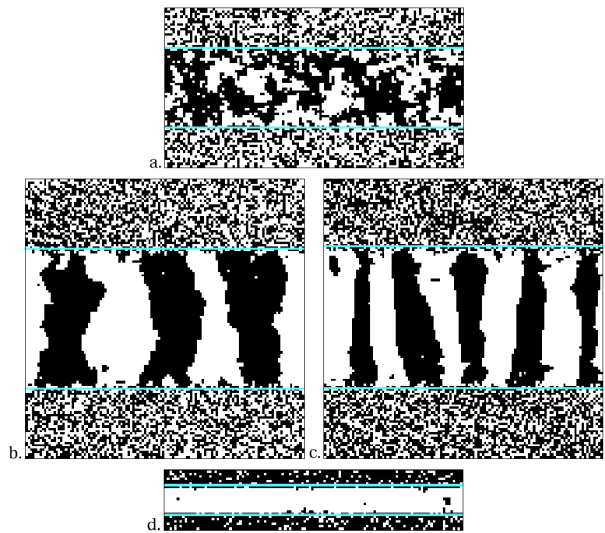}
\caption{\label{fig2} (Color online)
Typical steady state configurations, showing (a) the disordered phase, (b) and (c) two modulated phases with different wave numbers, and
(d) the fully ordered phase, all encountered when changing the temperature $T_0$ of the cold sector and/or the aspect ratio of the system.
The width of the system is always $2W = 120$. The system parameters are:
(a) $T_0=2.4$ and $L = 32$, (b) $T_0=1.0$ and $L=60$, (c) $T_0=0.4$ and 
$L=60$, and (d) $T_0=1.0$ and $L=12$. The gray (cyan online) lines indicate the boundaries between the two sectors.
For the purpose of showing both boundaries we have shifted the configurations by $L/2$ in the $y$-direction.
}
\end{figure}

In order to get a first impression of the complexity of the stationary states encountered in our system,
we show in Fig.\ \ref{fig2} some steady state configurations for various temperatures $T_0$ of the upper sector
as well as for various aspect ratios $L/W$, with $W = 60$. We readily identify different nonequilibrium phases.

Fig.\ \ref{fig2}a shows a typical configuration for the disordered phase that prevails when $T_0$ is larger than
the critical temperature $T_c = 2.269 \ldots$ of the two-dimensional Ising model (we set $J/k_B = 1$, with
$k_B$ being the Boltzmann constant). Due to the contact with 
the hot sector, correlated fluctuations inside the cold sector are taking place close to the interface
on much shorter length scales than for the corresponding equilibrium system. When moving further away from the
interface, the correlation length increases and approaches the correlation length of an equilibrium Ising system
at temperature $T_0$.

When $T_0$ is below $T_c$, phase separation sets in in the low temperature half. It is well known that the equilibrium Ising
lattice gas at half filling
separates into high and low density regions. In our system, however, a very
complicated ordered state emerges where the particles form stripes that span the cold sector in
the direction perpendicular to the interface, the different stripes being
separated by empty regions, see Fig. \ref{fig2}b and \ref{fig2}c. As shown in the figures for two cases
with 3 and 5 stripes, and as discussed 
in more detail below, the number of stripes changes
when the temperature $T_0$ and/or the aspect ratio changes, yielding a complex nonequilibrium phase diagram.

Finally, for very asymmetric shapes, i.e. for small values of $L/W$, another phase emerges where the particles 
and holes are 
surprisingly well separated. Fig. \ref{fig2}d shows a configuration where the particles are almost
completely sucked out of the low-temperature sector (due to the symmetry of the Ising Hamitonian,
one can of course also end up with the situation where particles are accumulating in this sector). All these particles end
up in the infinite temperature region which therefore has an extremely high particle density.

Let us pause here for a second to consider an {\it equilibrium} system that at first look might seem very similar
to the nonequilibrium system under investigation. In this system we set the coupling constants to be one in the upper
half of the system but zero in the lower half, with the coupling constants along bonds connecting
the two sub-systems being equal to one. The system itself is at a fixed temperature $T < T_c$ and the updates are done
using Kawasaki dynamics. As a result of the zero coupling constants in the lower sector, every proposed exchange will
be accepted in that sector. The main difference to our model can be found at the interface separating the sectors: for the equilibrium model
the exchanges across the interface fulfill detailed balance, whereas detailed balance is broken in our case. As a result
the upper sector of the equlibrium system simply phase separates, whereas the lower sector remains disordered, and none
of the intriguing non-equilibrium features discussed in this paper show up.
Especially, if we start for the equilibrium system with a phase separated upper sector and a disordered lower sector, 
phase separation will simply persist, as there is no mechanism in the equilibrium case that would allow to break up this
well ordered large domains. Consequently, modulated phases do not show up in this equilibrium system.

In order to fully characterize the periodic structures observed in our
nonequilibrium system we analyze the vertically resolved structure factor,
i.e. the norm squared of the Fourier transform of $S_{x,y}$ in the $x$-direction for fixed values of $y$
\footnote{A more complete information is contained in the two-dimensional structure factor,
but as we are interested here in the study of the modulated structures that form in $x$-direction,
we restrict ourselves to the Fourier transform in that direction.}:
\begin{equation}
      \label{eq:FT}
F_y(k) = \left| \frac{1}{2W}\sum_{x=1}^{2W} \, S_{x,y} \, e^{i k x} \right|^2~.
\end{equation}
For the sake of obtaining good statistics, we perform both a time and ensemble average. After reaching the steady state
we sample the quantity for another million time steps, repeating this procedure typically ten times.
In order to reduce the noise in the data we
found it useful to average our quantity over the middle third of the upper sector. It 
is this averaged quantity
\begin{equation}
      \label{eq:FT2}
F_{ave}(k) = \frac{3}{L} \sum\limits_{y=4L/3}^{5L/3} \langle F_y(k) \rangle ~,
\end{equation}
where $\langle \cdots \rangle$ indicates both the time and ensemble averages, that we use in order to construct
the nonequilibrium phase diagram discussed below.

\begin{figure}[h]
\centerline{ \includegraphics[width=0.90\columnwidth]{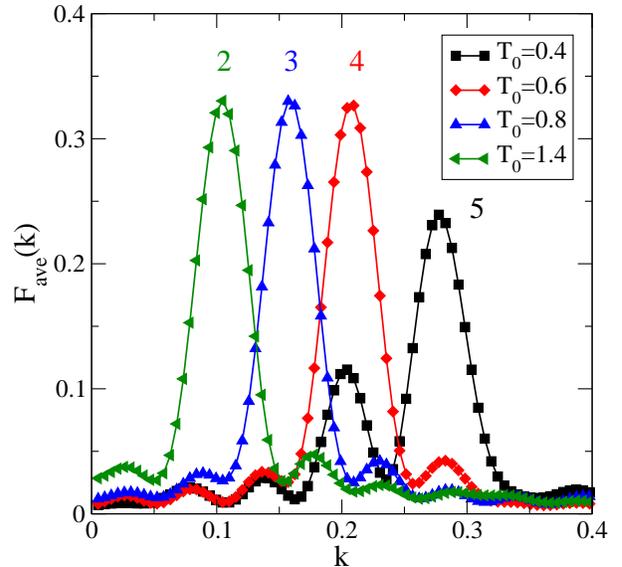} }
\caption{\label{fig3} (Color online) Averaged structure factor $F_{ave}(k)$ for a system with $W = L =60$,
at various temperatures $T_0$ of the cold sector. The peaks show up for well defined wave numbers $k$ that correspond 
to the number of stripes in the system.}
\end{figure}

The temperature dependence of the averaged structure factor $F_{ave}(k)$ for the system with $W =  L =60$  
is discussed in Fig. \ref{fig3}. In most of the cases shown in that figure  $F_{ave}(k)$ exhibits a single pronounced maximum
at some value $k = k_{max}$,
as for example for the temperature $T_0 = 1.4$, corresponding to a stable phase characterized by stripes
of width $\pi/k_{max}$. Thus, for $T_0 = 1.4$ we have 2 vertical stripes composed of particles with 
an average horizontal size of 30 lattice sites,
separated by stripes of the same size that are formed by holes. We will in the following characterize the different 
phases by the number of particle stripes.

For $T_0=0.4$ the average structure factor has two pronounced maxima, thus revealing the coexistence of two nonequilibrium phases
with different numbers of stripes (4 and 5 in our example). The presence of coexistence regions indicates that
the transitions between different phases are discontinuous.

\begin{figure} [h]
\centerline{ \includegraphics[width=0.95\columnwidth]{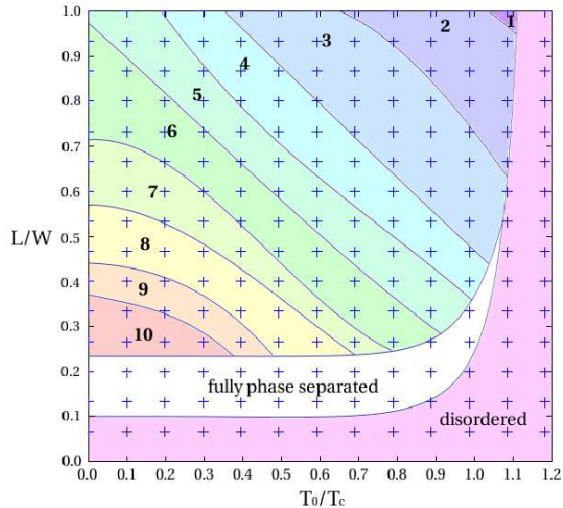} }
\caption{\label{fig4} (Color online) Phase diagram for a system of width $2W=120$ as a function of $T_0/T_c$ and $L/W$. 
Three different regions are readily
identified: the disordered phase at large values of $T_0/T_c$ and small values of $L/W$ without long range order,
the phase separated region where almost all particles can be found in one sector of the system, and the
modulated region where different periodic arrangements show up in the cold temperature
sector of the system. The crosses indicate the temperatures and system sizes at which the numerical
simulations were done. The lines separating the different phases result from cubic splines through
the midpoints between crosses that yield different phases.
}
\end{figure}

We use the information from the structure factor in order to construct the phase diagram shown in Fig.\ \ref{fig4}.
We can distinguish three different regions, corresponding to the different types of steady states illustrated
in Fig.\ \ref{fig2}: the disordered phase without long range order, the almost perfectly phase separated state where
an overwhelmingly large fraction of particles is sucked into one part of the system, and the modulated region with a multitude of 
striped phases. 
As inside the modulated region the transitions between the phases are discontinuous,
the lines shown in that part of the diagram stand for the larger coexistence regions that
separate different modulated phases. It is worth noting that in our finite systems
phases with strip lengths that are incommensurable 
with the lattice size are suppressed. In the infinite volume limit $L, W \longrightarrow \infty$, with
$L/W$ kept constant, the stripe width is expected to change continuously.

Studying systems with sizes ranging from $W = 30$ and $W=120$, we observe
the same three regions in the phase diagram.
The areas occupied by the disordered phase and by the phase separated region in the $T/T_c-L/W$ phase diagram
decrease when the lateral extension of the system increases, indicating that they will vanish in the
infinite volume limit.
Inside the modulated region the number of
stable phases increases with $W$. Thus for $W=30$ we only observe phases that contain up to 5 stripes,
whereas for $W=120$ we can identify phases with up to 20 stripes. All this indicates that the modulated
regions are not due to finite size effects and should therefore also persist in larger systems.
The apparent persistence of the modulated region for temperatures slightly above $T_c$,
the critical temperature of the infinite system, is
due to the well known shift of the (pseudo-)critical temperature in finite systems.

In order to understand the appearance of stripes in this system with conserved dynamics, we point out that a recent study \cite{Ple10}
showed the emergence of convection cells, driven by spontaneous symmetry breaking,
when the cold sector temperature $T_0$ is set below the equilibrium critical
temperature. In that work pinned boundary conditions at the $y$-boundaries were used in order to facilitate the measurement
of the convection cells through the vorticity $\omega$ and the stream function. For the periodic boundary conditions 
used in our work the time and ensemble averaged vorticity $\langle \omega \rangle$ is trivially zero. Non-trivial
insights into the emergence and persistence of convection cells, as well as of their relationship to the modulated
phases, can be gained by studying (we here go back to the language of magnetic spin systems) the spin-vorticity correlation
function $\langle S_{x',y'} \omega(x,y)\rangle$ where $\omega(x,y)$ is the vorticity related to the plaquette centered
at $(x+\frac{1}{2},y+\frac{1}{2})$. The vorticity is thereby measured through 
\begin{equation}
\omega(x,y) = j_y(x,y) + j_x(x,y+1) - j_y(x+1,y) - j_x(x,y)
\end{equation}
where the $j$'s are the net currents
across each bond around the plaquette \cite{Ple10}, i.e.\ $\omega(x,y) > 0$ for a counter-clockwise rotation
around the plaquette.

\begin{figure} [h]
\centerline{ \includegraphics[width=0.9\columnwidth]{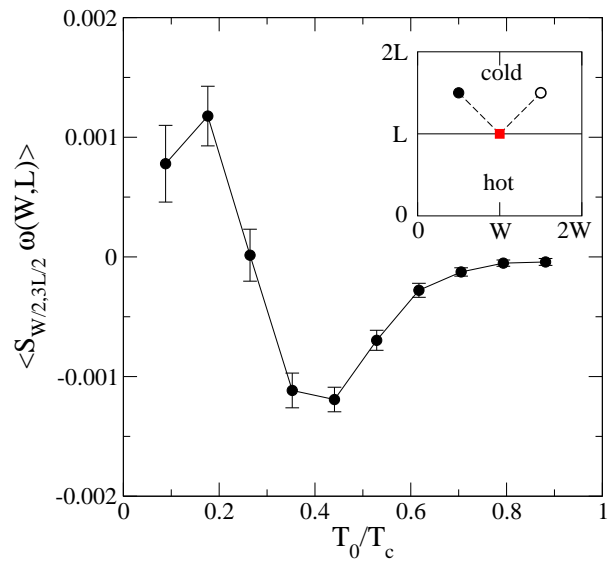} }
\caption{\label{fig5} (Color online) Temperature dependence of the
spin-vorticity correlation $\langle S_{W/2,3L/2} \, \omega(W,L) \rangle$
that relates the vorticity around a plaquette in the middle of the 
interface separating the hot and cold sectors to the spin $S_{W/2,3L/2}$ located inside
the cold sector. The system contains $2W\times 2 L$ sites, with $W = L = 60$.
The inset shows the positions of the plaquette (square) and of the spin $S_{W/2,3L/2}$ (full circle)
in the sample. The correlation $\langle S_{3W/2,3L/2} \, \omega(W,L) \rangle$
with the spin $S_{3W/2,3L/2}$, shown by the open circle in the inset,
only differs from $\langle S_{W/2,3L/2} \, \omega(W,L) \rangle$ by the sign.
These data result both from a time average and an ensemble average over 30 
different realizations of the noise.
}
\end{figure}

Fig.\ \ref{fig5} shows the temperature dependence of
$\langle S_{W/2,3L/2} \, \omega(W,L) \rangle$ that relates the
vorticity around a plaquette in the middle of the interface between the hot and cold sectors to a spin located in the middle
of the cold sector. We immediately note that the spin-vorticity correlation is non-vanishing, thereby revealing
the presence of convection cells when $T_0$ is brought below the critical temperature. Due to the symmetry
between the locations of the two spins $S_{W/2,3L/2}$ and $S_{3W/2,3L/2}$
the two correlation functions $\langle S_{W/2,3L/2} \, \omega(W,L) \rangle$ and
$\langle S_{3W/2,3L/2} \, \omega(W,L) \rangle$ have the same magnitude
but different signs. The spin-vorticity correlation also shows features that are due to the
transitions between modulated phases. 
\footnote{We can restrict the following discussion to configurations where the plaquette is located at a stripe
boundary, as only these configurations yield a non-vanishing vorticity.}
Thus the change in sign around $T_0 \approx 0.26~T_c$
is readily understood by the phase sequence 3 stripes $-$ 4 stripes $-$ 5 stripes when
lowering $T_0$. Writing the 3 stripes phase as $+-+|-+-$, where the vertical line indicates the position of the plaquette along
the $x$ axis whereas a $+$ ($-$) sign corresponds to a $+$ ($-$) stripe, 
we expect $\omega > 0$ around our plaquette, due to the sequence $+|-$ \cite{Ple10}. 
The spin $S_{W/2,3L/2}$ being located in a $-$ region, this then yields 
a negative sign for $\langle S_{W/2,3L/2} \, \omega(W,L) \rangle$.
This is different for the case of 5 stripes, $+-+-+|-+-+-$, that yields again
a counter-clockwise rotation with $\omega > 0$, but now with the spin $S_{W/2,3L/2}$ in a $+$ region.
These two phases are separated by the phase with four stripes, $+-+-|+-+-$, where our spin is located at the boundary
between $+$ and $-$ stripes, yielding a strong suppression of the corresponding spin-vorticity correlations.

These results indicate that it is the presence of convection cells that ultimately leads to the formation of the modulated
phases in our nonequilibrium system. Whereas the corresponding equilibrium system separates below the critical temperature
into two domains with high and low densities, these large domains are broken up in our system due to the subtle interplay
of phase separation and convection facilitated by the presence of the high temperature region. An analytical description
of this complicated situation remains a challenge and is left for future studies.

We expect this type of behavior to be generic for systems undergoing phase separation where different sectors 
are coupled to different thermal baths and where detailed balance is broken locally.
In this situation spontaneous symmetry breaking leads to the
emergence of convection cells that favor the formation of modulated nonequilibrium phases. We have restricted
ourselves in this communication
to the situation where the hot sector is at infinite temperatures. A detailed discussion of the situation where that
sector is at a finite temperature $T' \geq T_c$ will be published elsewhere.

In summary, our study reveals a complex nonequilibrium phase diagram for an Ising lattice gas where 
different sectors of the system are in contact with thermal baths with markedly different temperatures. In case that
one sector is at a temperature above $T_c$ whereas the other is at a temperature below $T_c$, two mechanisms
set in, namely spontaneous symmetry breaking and convection, that together lead to the formation of remarkable
nonequilibrium modulated
stationary states.

\acknowledgments
We thank Beate Schmittmann and Royce Zia for many enlightening discussions and Uwe T\"{a}uber for a critical
reading of the manuscript. This work was supported by the US National
Science Foundation through DMR-0904999.

\end{document}